\documentclass[preprint]{aa}
\usepackage{natbib}
\usepackage{amsmath,amssymb}

\begin{document}

\title{A physical model for the magnetosphere of Uranus at solstice time}

\titlerunning{The magnetosphere of Uranus at solstice}

\author{Filippo Pantellini}
\institute{LESIA, Observatoire de Paris, Universit\'e PSL, CNRS, Sorbonne Universit\'e, Universit\'e de Paris, 5 place Jules Janssen, 92195 Meudon, France}


\abstract{
Uranus is the only planet in the Solar System whose rotation axis and orbital plane are nearly parallel to each other. Uranus is also the planet with the largest angle between the rotation axis and the direction of its magnetic dipole (roughly $59^\circ$). Consequently, the shape and structure of its magnetospheric tail is very different to those of all other planets in whichever season one may consider. The only in-situ measurements were obtained in January 1986 during a flyby of the Voyager II spacecraft. At that date,  Uranus was near solstice time, but unfortunately the data collected by the spacecraft were much too sparse to allow for a clear view of the structure  and dynamics of its extended magnetospheric tail. Later numerical simulations revealed that the magnetic tail of Uranus at solstice time  is helically shaped with a characteristic pitch of the order of 1000 planetary radii.}{
We aim to propose a magnetohydrodynamic model for the magnetic tail of Uranus at  solstice time. 
}{
We constructed 
our model based on a symmetrised version of the Uranian system by assuming an exact alignment of the solar wind and the planetary rotation axis and an angle of $90^\circ$ between the planetary magnetic dipole and the rotation axis. We do also postulate that the impinging solar wind is steady  and unmagnetised, which implies that the magnetosphere is quasi-steady in the rotating planetary frame and that there is no magnetic reconnection at the magnetopause.   
}{
One of the main conclusions is that all magnetic field lines forming the extended magnetic tail follow the same qualitative evolution from the time of their emergence through the planet's surface and the time of their late evolution after having been stretched and twisted several times downstream of the planet. In the planetary frame, these field lines move on magnetic surfaces that wind up to form a tornado-shaped vortex with two foot points on the planet (one in each magnetic hemisphere). The centre of the vortex (the eye of the tornado) is a simple double helix with a helical pitch 
(along the symmetry axis $z$) $\lambda=\tau[v_z+B_z/(\mu_0\rho)^{1/2}],$ where $\tau$ is the rotation period of the planet, $\mu_0$ the permeability of vacuum, $\rho$ the mass density, $v_z$ the fluid velocity, and $B_z$ the magnetic field where all quantities have to be evaluated locally at the centre of the vortex. In summary, in the planetary frame, the motion of a typical magnetic field of the extended Uranian magnetic tail is a vortical motion, which asymptotically converges towards the single double helix, regardless of the line's emergence point on the planetary surface.       
}{}


\maketitle

\section{Introduction}

The magnetosphere of Uranus was traversed by the Voyager II spacecraft on January 1986. Most of the planet's parameters such as the rotation period, the internal magnetic dipole strength, and the orientation of the associated axis were discovered or refined on that occasion \citep{Ness_etal_1986, Desch_etal_1986,Bagenal_1992, Bagenal_2013, Richardson_Smith_2003}. Since then,   
we have also learned that the solar wind impacting the magnetosphere has a high Mach number and low $\beta$ (the thermal pressure to magnetic pressure ratio), with a typical sonic Mach number in the range of 20 to 30 and $\beta$ in the range of 0.1 to 0.2. These values imply wind velocities largely in excess of the fast magnetosonic speed with typical fast Mach numbers in the 6 to 12 range. Voyager's measurements also showed that  
the planet's internal magnetic dipole intensity is high enough to counter the solar wind's dynamic pressure. Uranus is thus surrounded by a  
magnetopause (the boundary between the solar wind flow and the regions dominated by the planetary magnetic field) and by a bow shock located upstream of the magnetopause. In the sub-solar direction, the magnetopause and the bow shock are located  
at a distance of $\sim 25 R_{\rm U}$ and $\sim 30 R_{\rm U}$ upstream of the planet's centre ($R_{\rm U}$ is the radius of Uranus).  
Numerical simulations show that the extended nightside Uranian magnetosphere is a particularly complex one compared to the other planetary magnetospheres in the Solar System. Such a complexity is due to both the large angle of $59^\circ$ between the planet's rotation axis and a rotation period fast enough (sidereal period: 17.24h) to be comparable to the magnetosphere's relaxation time. 
The magnetosphere of Uranus is thus highly variable on a daily timescale.  In addition, the planet's rotation axis being nearly parallel with respect to the  orbital plane also makes the magnetosphere highly variable on a seasonal timescale (with an orbital period of 84 years). For comparison, Saturn's rotation and magnetic axis are aligned \citep[see e.g.][Fig. 2]{Cowley_2013}, so under constant wind conditions its magnetosphere is stationary, which is an enormous difference with respect to Uranus. 
Unfortunately, in-situ measurements of the Uranian magnetosphere are very sparse. Voyager II, the only spacecraft having approached Uranus, only explored a small portion of the planet's magnetic tail of the order of $\sim 60 R_{\rm U}$, which is a scale much shorter than the characteristic spatial oscillations of $10^3 R_{\rm U}$ observed in the pioneering simulations by \cite{Toth_etal_2004}. Since that time, only a relatively small number of simulations have been published.  The reasons are likely the lack of new observational data for the simulations to be confronted with and the exceptionally large simulation domain required to capture the oscillations of the magnetic tail. As for \cite{Toth_etal_2004}, all subsequent simulations are also based on single or multi-fluid equations for equinox 
configuration \citep{Cao_Paty_2017,Griton_etal_2018} and solstice configuration \citep{Cao_Paty_2017, Griton_Pantellini_2020, Lai_Kiang_2020}. It is however worth mentioning that only in  \cite{Toth_etal_2004} is the simulation domain large enough to allow for a fully developed magnetic tail. This is also the case in the simulations by \cite{Griton_etal_2018} and \cite{Griton_Pantellini_2020}, where the authors opted to accelerate the planet's rotation in order to reduce the spatial scale of the tail oscillation and thus the size of the simulation domain.  \cite{Cao_Paty_2017} and \cite{Lai_Kiang_2020} focused on the dayside structure of the magnetosphere, which allows for a small simulation domain but excludes the possibility to observe the oscillations in the tail of the simulation, which is a central aspect of the physical model we present in this study. 

The physical model presented here is restricted to the case of Uranus at solstice. In order to reduce complexity, while keeping all the fundamental ingredients that make the specificity of Uranus, we symmetrised the solstice configuration. First, we increased the angle between the magnetic dipole and the rotation axis from $59$ to $90^\circ$. Second, we aligned the planet's rotation axis and the solar wind direction, which are in reality separated by $7^\circ$.  Finally, we postulate that the solar wind is not magnetised (as in the \citet{Toth_etal_2004} simulations) so that no magnetic reconnection is allowed in the model. Reconnection is be briefly discussed at the end of this paper. A more in-depth discussion of the subject can be found in the work by \cite{Masters_2014}. 

The paper is organised as follows. In Sections \ref{sec_configuration} and \ref{sec_assumptions}, we comment on the symmetries and the physical assumptions of the model. In Sections \ref{sec_path} and \ref{sec_wavelength}, we discuss the time evolution and the twisting of a typical magnetic field line in the  magnetospheric tail. In Section \ref{sec_current}, we give a crude estimate of the pole-to-pole electric current induced by the planet's rotation. In Sections \ref{sec_limits} and \ref{sec_speculations}, we briefly comment on the model's limits and speculate on the generic case (with no symmetries). Conclusions are given in Section \ref{sec_conclusions}.

\section{A simplified configuration of the problem \label{sec_configuration}}

\begin{figure}[h]
        \center{\rotatebox{0}{\resizebox{0.45\textwidth}{!}{\includegraphics{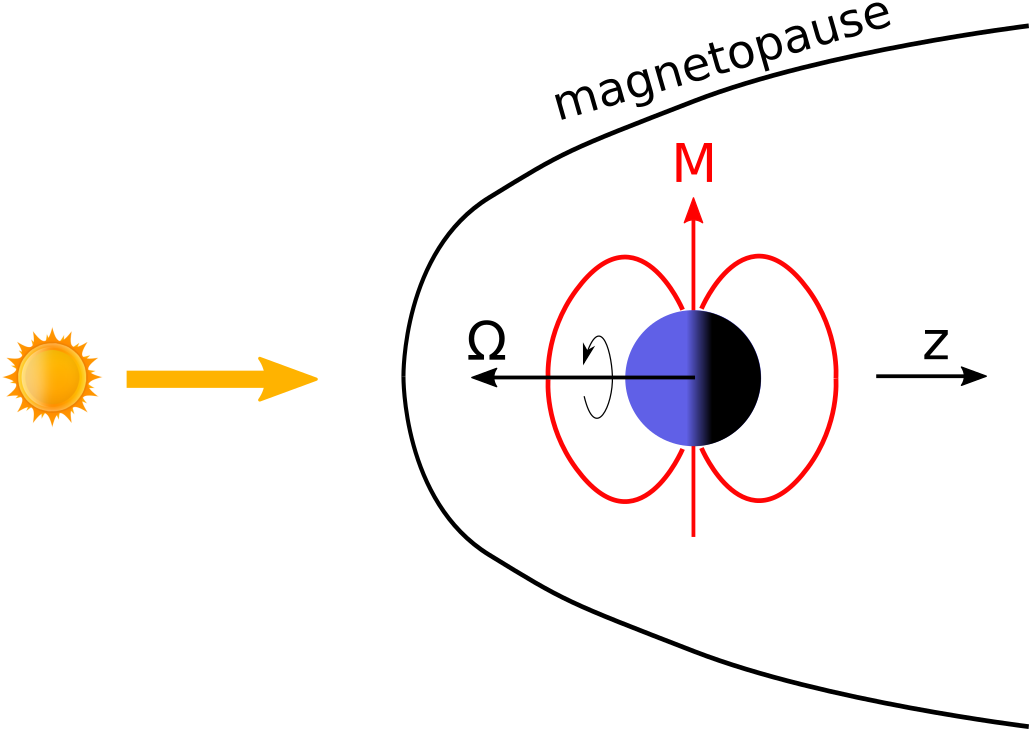}
        }}}
        \caption{Symmetrised configuration of Uranus at solstice used in our physical model. The rotation axis and solar wind flow are both aligned on the $z$ axis. $\Omega$ is the planet's angular velocity. The planetary field is a dipole of strength $M$ located at the centre of the planet and oriented perpendicularly to the $z$ axis.} 
        \label{fig_1}
\end{figure}

In order to reduce the complexity of the real system and facilitate insight, we assume that the planet's rotation axis points in the direction opposite to the solar wind direction. We also postulate that the magnetic dipole of the planet is centred on the planet and oriented perpendicularly to the rotation axis, as illustrated on Fig. \ref{fig_1}.

In addition to assuming a stationary solar wind, we also assume that it does not carry any magnetic field. The wind is therefore rotationally invariant with respect to the z axis. Also, in agreement with Voyager's measurement, the wind is assumed to be supersonic and the planetary's internal field strong enough to support an anti-sunward-oriented magnetopause as shown in Fig. \ref{fig_1}. We add that, during its flyby, Voyager II saw the magnetic field intensity rise by one order of magnitude at the magnetopause, providing an observational justification for neglecting the magnetic field carried by the solar wind \cite[see][]{Ness_etal_1986}. As there are no field lines in the wind, the magnetopause can also be identified as the magnetic surface traced by the outermost magnetic field lines connected to the planet. 

\subsection*{Symmetries of the simplified configuration \label{sec_symmetries}}

The system depicted in Fig. \ref{fig_1} has many interesting symmetries.  First, since the wind direction and the planet's rotation axis are aligned and since the former is rotationally invariant, the system can be assumed to be  time stationary in the rotating frame, at least on timescales of the order of the rotation period. Second, the system is invariant under successive application of the following actions: (1) a specular reflection with respect to a plane containing the $z$ axis followed by (2) a specular reflection with respect to another plane perpendicular to the first one and also sharing the $z$ axis, followed by (3) a reversal of all charges in the system. We note that action (3) corresponds to a symmetry of the magnetohydrodynamic (MHD) system of equation, which does not hold at small (kinetic) scales. In practice, the action of reversing the charges in an MHD system corresponds to the action of reversing the sign of all currents and magnetic fields in the system.       
\begin{figure}[h]
        \center{\rotatebox{0}{\resizebox{0.3\textwidth}{!}{\includegraphics{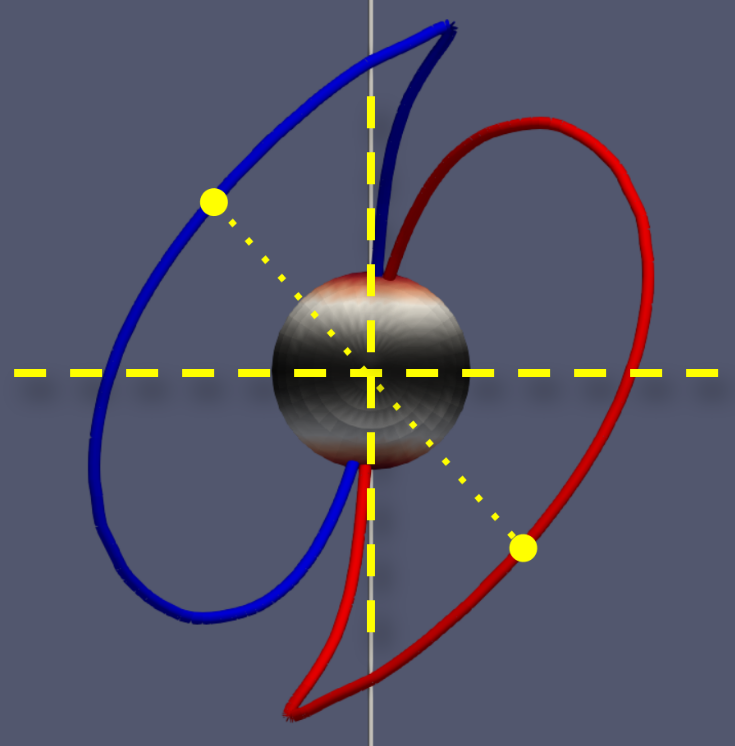}
        }}}
        \caption{Example of symmetries of system in Figure \ref{fig_1}. System is viewed looking to the planet along the $z$ axis. Two mutually conjugate magnetic field lines are shown, as well as two conjugate points (represented by the two large yellow dots).  
        \label{fig_2}}
\end{figure}
Figure \ref{fig_2}, extracted from a 3D MHD simulation respecting the symmetries of the configuration of Figure \ref{fig_1}, shows two field lines that are mutually symmetric with respect to application of the three mentioned symmetry actions. For the remainder of the paper, we call such pairs of field lines 'conjugate field lines', and the points with reversed $x$ and $y$ coordinates are referred to as 'conjugate points' (large yellow points in the figure).    
We do not list all consequences of the system's symmetries. We do however note that one important consequence is that only the $z$ component of the fluid velocity can be non-zero on the $z$ axis, which also implies that magnetic field lines are not allowed to cross the $z$ axis. Another consequence is that on conjugate points, the $z$ components of both the current and the magnetic field are reversed, while the perpendicular components (with respect to the $z$ axis) are equal. The exact contrary holds for the fluid velocity.    

\section{Model assumptions \label{sec_assumptions}}

Besides the symmetries mentioned in the previous section, we have to make some assumptions about the boundaries. These are the magnetopause and the planetary surface. Concerning the magnetopause, we assume it to be an impermeable surface. No plasma from the wind can penetrate through the surface, which is coherent with the assumption of no magnetic field in the wind. On the other hand, the planetary surface is the only source or sink of magnetic flux. As already mentioned, we assumed a dipolar intrinsic planetary field, which we placed  at the planet's centre, implying that field lines emerge at the planet's magnetic equator. The plasma is governed by the equations of ideal MHD everywhere except at the planet's surface, where a finite conductivity is assumed such that the foot points of the magnetic field lines are allowed to move across the surface. Even though the detailed structure of the magnetosphere may depend on the boundary conditions, we assume that the latter are not critical in shaping the distant tail structure and choose not to specify these boundaries further. 
As for most plasmas in the Solar System, the interaction of the solar wind with a planetary magnetosphere can be treated in the weakly relativistic limit. We therefore assume that the plasma in our model is governed by the equations of  magnetohydrodynamic limited to first order in $v/c$, where $v$ is a characteristic velocity (fluid velocity or phase velocity), and $c$ the speed of light. Under such circumstances when changing from one frame of reference to another (for example, from the planetary rotating frame to the inertial frame), velocities transform according to the Galilean transformation rules. All other quantities (density, pressure, and magnetic field) are invariant. Finally, on Uranus, the escape velocity from the planet's surface is small compared to the Alfvén speed. We therefore ignored gravity.

\section{Trajectory of a magnetic field line\label{sec_path}}

\begin{figure}[h]
        \center{\rotatebox{0}{\resizebox{0.45\textwidth}{!}{\includegraphics{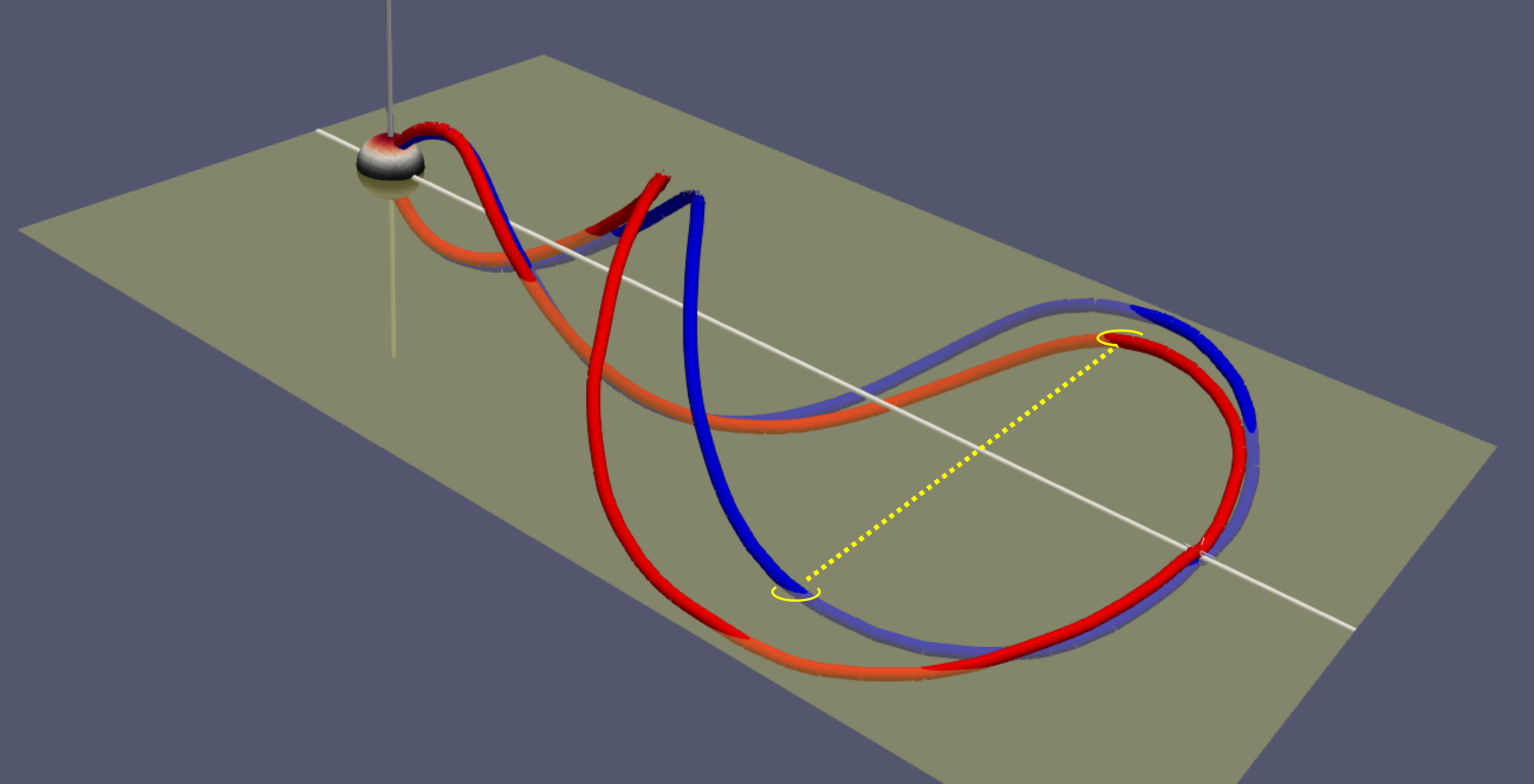}
        }}}
        \caption{Points corresponding to the intersection of two conjugate magnetic field lines with any plane containing the $z$ axis are disposed symmetrically on each side of the axis.} 
        \label{fig_3}
\end{figure}

\begin{figure}[h]
        \center{\rotatebox{0}{\resizebox{0.45\textwidth}{!}{\includegraphics{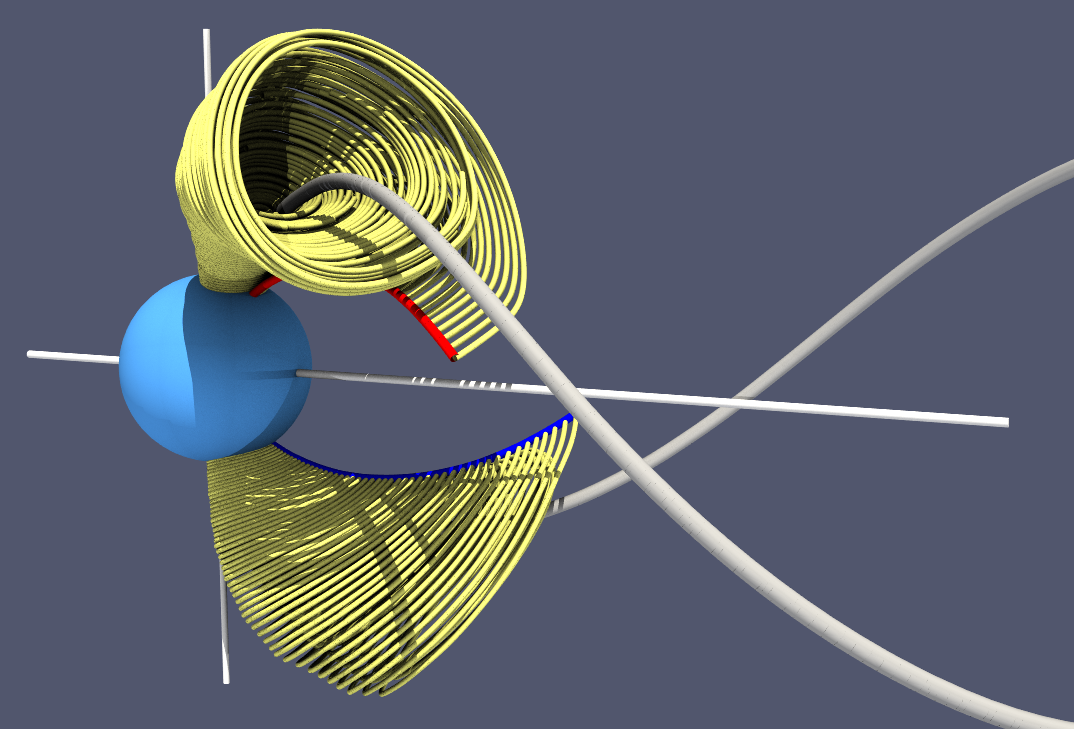}
        }}}
        \caption{Planetary (rotating) frame. Magnetic surfaces described by the time evolution of two conjugate magnetic field lines (red and blue). Only a portion of the surface described by the future evolution of the two field lines is shown. We anticipate that the thick grey tubes show the end positions of the field lines.} 
        \label{fig_4}
\end{figure}

In a three-dimensional system, magnetic field lines can form complex and interlaced structures. The difficulty of apprehending a complex three dimensional magnetic structure is even greater in a case where the structure is a time-dependent one. As already mentioned, the model, based on the symmetries of Figure \ref{fig_1}, can be considered to be time stationary in the rotating frame. Thus, in the following, unless specified otherwise, we place ourself in the rotating frame. The downside of choosing the rotating frame instead of the inertial frame is that in addition to the standard forces, which are the pressure gradient and the Lorentz force, we have to take into account the Coriolis and the centrifugal forces in the momentum equations \cite[see e.g.][]{Toth_etal_2004}. In order to further facilitate the investigation of a complex three-dimensional structure, it is sometimes instructive to trace the motion of the intersection of the magnetic field lines with a given plane. Figure \ref{fig_3} illustrates the case of the intersection of two conjugate field lines with the magnetic equatorial plane. We note that due to the system's symmetries, the intersection points of the conjugate field lines are always equidistant from the $z$ axis, for any arbitrary plane that contains the $z$ axis. The description of the time evolution of the intersecting points of the magnetic field lines with the equatorial plane is one of the main objectives of the present study (see Section \ref{sec_path_equator}). 
However, before proceeding to the formulation of a physical model of the simplified version of the magnetosphere of Uranus at solstice, it is worth mentioning another of its properties by noting that field lines successively emerging at a given position on the planet's surface do all follow the exact same path. Obviously, all these field lines together define a magnetic surface. An example of a portion of such a magnetic surface described by the  time evolution of two conjugate field lines is shown in Figure \ref{fig_4}.  
Figure \ref{fig_4} illustrates a central aspect of the system: field lines move towards an asymptotic position (the grey tubes located inside the magnetic surfaces). From the figure, one may already deduce that the trajectory of the intersecting point of the field line with an arbitrary plane is a spiral with the asymptotic position at its centre.

\subsection{Trajectory on the planetary surface and in the (magnetic) equatorial plane \label{sec_path_equator}}

\begin{figure*}[h!]
        \center{\rotatebox{0}{\resizebox{0.9\textwidth}{!}{\includegraphics{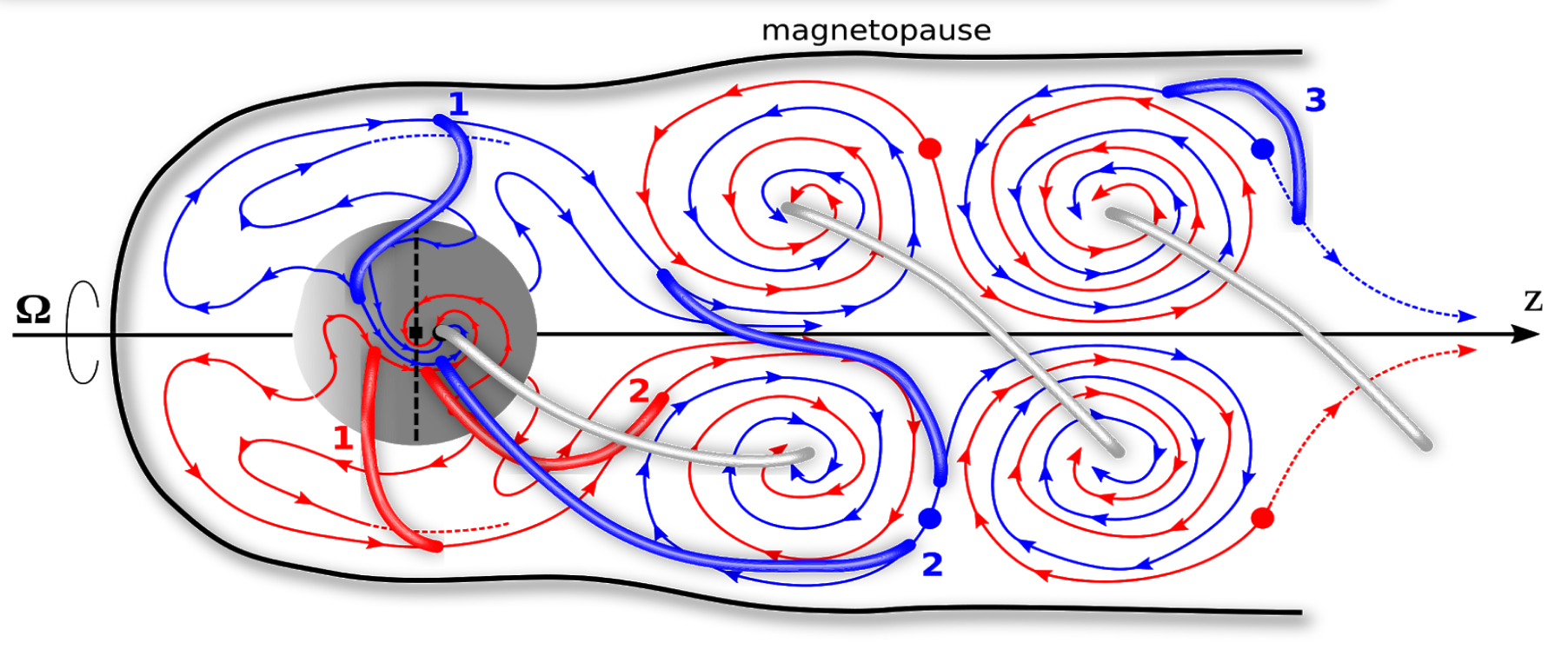}
        }}}
        \caption{Polar view in the rotating frame. Shown are the  trajectory of the foot points of selected field lines in the ionosphere (one hemisphere only), as well as the trajectory of the field lines' intersection points with the equatorial plane. Blue and red correspond to mutually conjugate lines. The grey tubes represent the asymptotic position of all field lines.} 
        \label{fig_5}
\end{figure*}

Having assembled all the basic ingredients, and based on results from numerical simulations, we are now in a position to propose a coherent physical model of a Uranus-at-solstice-type magnetosphere. We start, in Figure \ref{fig_5}, with the presentation of a qualitative sketch of the path of the field line crossing point in the equatorial plane and the path of their foot points on the planetary surface. We note that, as a consequence of the difference between the velocity of a fluid parcel and the velocity of a virtual point, such as the point corresponding to the intersection of a field line with a plane, the pattern described by the projected fluid velocity can be very different from the pattern described by the motion of a virtual point, as in Figure \ref{fig_5}. 
This is namely the case for the multiple spiralling trajectories in the equatorial plane that are not seen when tracing the projection of the fluid velocity. The reason is that in some parts of the equatorial plane, for example near position 2 in the figure, field lines are oriented nearly parallel with respect to the plane, in which case the velocity of the intersection point and the fluid velocity differ substantially. On the other hand, the projected fluid velocity and the velocity of the intersection point coincide when the field line is perpendicular to the plane. 
Of course, the path described by a field line depends on the basic parameters of the problem (rotation period of the planet, solar-wind dynamic pressure, planetary field strength, etc.), but also on the position of its point of emergence on the planetary surface. However, from the topological point of view, all field lines behave as described in Figure \ref{fig_5} as long as they do not go back to the planet, an eventuality not shown on the figure. 

We now comment more thoroughly on Figure \ref{fig_5}, where 
three successive positions of a typical magnetic field line (in this case the blue line) are labelled with numbers. At position 1, the foot point of the line on the planetary surface and the crossing point of the field line on the equatorial plane have already moved away from their emerging point at the  magnetic equator. Simply as a consequence of its shape, which makes the emerging field line parallel to the planetary surface, the foot point of an emerging dipolar field line moves towards the pole. Obviously, the poleward motion of the foot point does not simply follow a meridian line as its motion is guided by various forces (pressure gradient, Lorentz, Coriolis, and centrifugal) and by the properties of the boundary. We note that a similar, but not symmetric poleward motion affects the second foot point located below the equatorial plane. However, recalling the symmetries of the system (see Figure \ref{fig_2}), we know that the shape of a given field line in the hidden  hemisphere is equal to the shape of the conjugate field line in the visible hemisphere. Thus, in Figure \ref{fig_5}, the hidden part of the blue line can be obtained by applying the transformations $x \rightarrow -x$ and $y \rightarrow -y$ to the red line. 

In order to keep things as simple as possible, even though emerged field lines may disappear again through the planetary surface, we only considered the case of field lines not disappearing through the planet's surface. The foot point of these lines must then converge towards some asymptotic position near the nominal magnetic pole. As all foot points move along spirals that also represent the trace of the corresponding magnetic surfaces (which do not cross by definition), it follows that the field lines must converge towards a common position represented by the black square in Figure \ref{fig_5}. 
Because of the tailward bending of the field lines, the asymptotic position is  expected to be located tailward of the nominal magnetic pole. Figure \ref{fig_5} suggests that an asymmetric polar vortex in the ionospheric plasma circulation has to be expected on Uranus. Distant observations of auroras using the Hubble Space Telescope are become possible and may be used as a tool to get access to the magnetic structure near the magnetic poles \cite{Lamy_etal_2012,Lamy_etal_2017}. Unfortunately, besides the fact that observed signals are quite faint and not sufficiently sharp to offer further help, no distant observations have ever been made at solstice. The next opportunity is in 2028, 42 years after the last occurrence in 1986.   

We now take inspiration from past MHD simulations \cite[see e.g.][]{Toth_etal_2004,Griton_etal_2018} to understand the more profound reason for the formation of the vortices in Figure \ref{fig_5}. Indeed, one of the  basic outcomes of these simulations is that the tailward extending magnetic field lines wind helically around the $z$ axis. This apparently trivial and intuitive result is an important constraint for our physical model. Looking at Figure \ref{fig_5}, we note that the only possible way to make the blue line in position 1 wind around the $z$ axis is to stretch the visible part of the line over both the pole and the $z$ axis until, guided by the magnetopause, it crosses the equatorial plane near the point labelled 2. Once the return point of the field line has disappeared below the equatorial plane, the line  continues to be stretched and twisted by $180^\circ$ before emerging at position 3, and so on. One should note that all these movements are constrained by the prohibition of the field line to intersect the the $z$ axis. For completeness, the path of the conjugate line (in red) is also shown in Figure \ref{fig_5}. 

As pointed out earlier (see discussion on Figure \ref{fig_4}), the intersecting points of a field line with the equatorial plane are seen to move on a spiral converging towards a final position given by the intersecting points of the light grey tubes in Figure \ref{fig_5}. Thus, the two (red and blue) field lines labelled 2 will gradually move towards the leftmost grey tube in the figure, the upper strand of the blue line labelled 2 will move towards the central grey tube, and the blue line labelled 3 will move towards the rightmost grey tube. We note that the polarity of the field lines converging towards the position of the central tube (which is connected to the magnetic pole below the equatorial plane) is opposite with respect to the polarity of the field lines converging towards the other visible portions of the grey tube (this portion being connected to the visible pole).

\subsection{Trajectory in a plane perpendicular to the z axis\label{sec_perp_z}}

\begin{figure}[h]
        \center{\rotatebox{0}{\resizebox{0.45\textwidth}{!}{\includegraphics{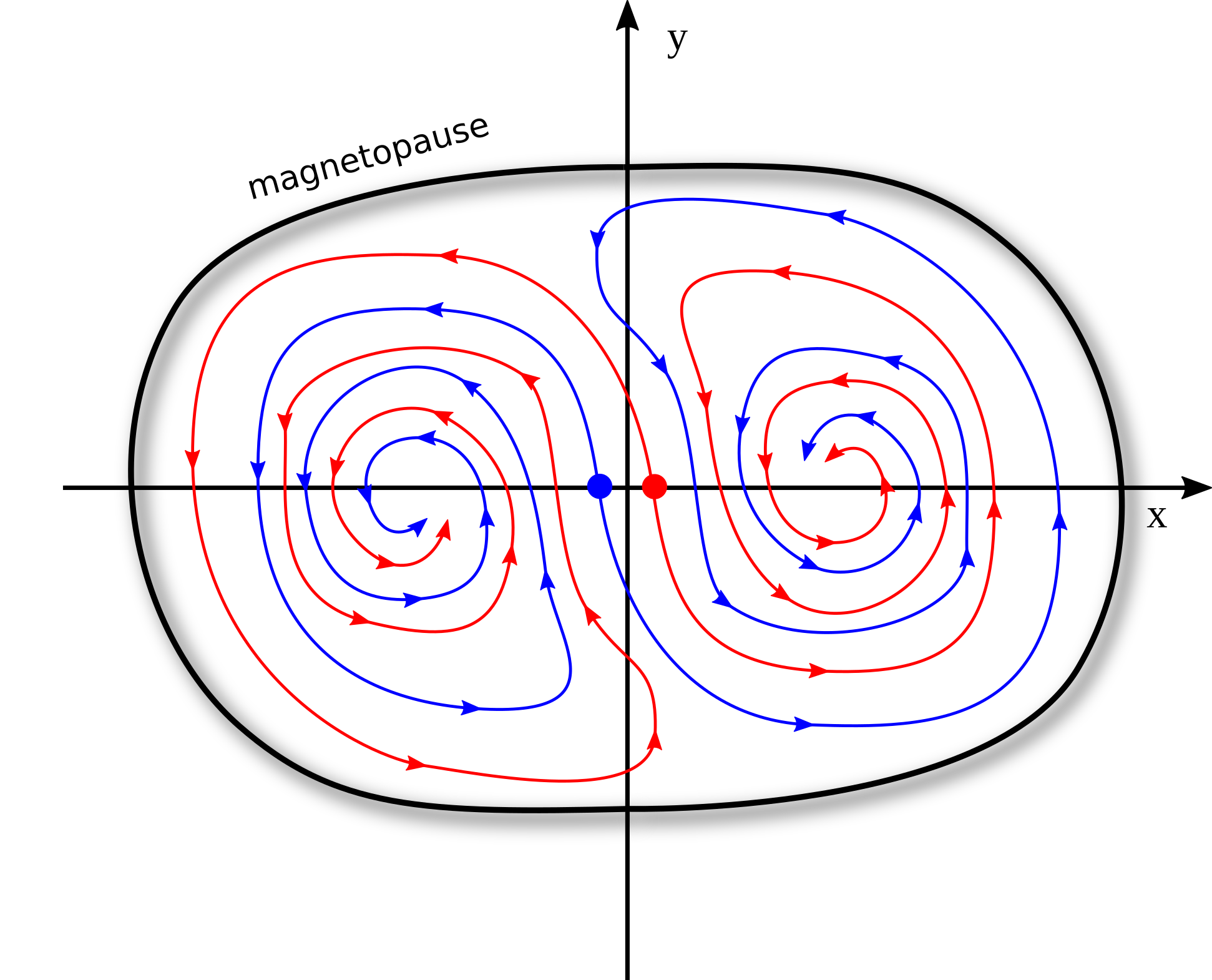}
        }}}
        \caption{Planetary rotating frame. Trajectories of the intersection point of two conjugate magnetic field lines with a plane perpendicular to the $z$ axis located downstream of the planet. Colours distinguish the trajectory of two mutually conjugate lines. The large blue and red dots indicate the position where the two conjugate lines cross the plane in the first place.} 
        \label{fig_6}
\end{figure}
Figure \ref{fig_6} shows the trajectory of two conjugate field lines in a plane perpendicular to the $z$ axis approximately positioned at the $z$ location and corresponding to the centre of a couple of vortices in Figure \ref{fig_5}. The two points near the origin indicate the position where the two field lines appear first, that is, at the time when the field lines' loop-return points reach the plane. As in the equatorial plane, after emergence, the two intersecting points of each of the two conjugate field lines (the red and the blue line) move on spiral shaped trajectories towards the asymptotic position at the centre of the two vortices. Again, it may be worth pointing out that the polarity (i.e. the $z$ component of the magnetic field) is opposite in the two vortices.          
We now make a few more comments on Figure \ref{fig_6}. First, we note that at the position where the field lines appear, the magnetic field is necessarily oriented perpendicularly to the $z$ axis, so the field line velocity and the plasma velocity in the $z$ direction must be equal at this particular position. 
One of the macroscopic consequences of this is that the whole magnetic tail structure as seen from an inertial frame is seen to propagate downstream of the planet at exactly the plasma speed at these locations. Second, except in the region of acceleration near the planet, the structure of Figure \ref{fig_6} is representative for all positions along the $z$ axis modulo a rotation about the $z$ axis.

\section{Wavelength of the magnetic tail \label{sec_wavelength}}

\begin{figure}
        \center{\rotatebox{0}{\resizebox{0.45\textwidth}{!}{\includegraphics{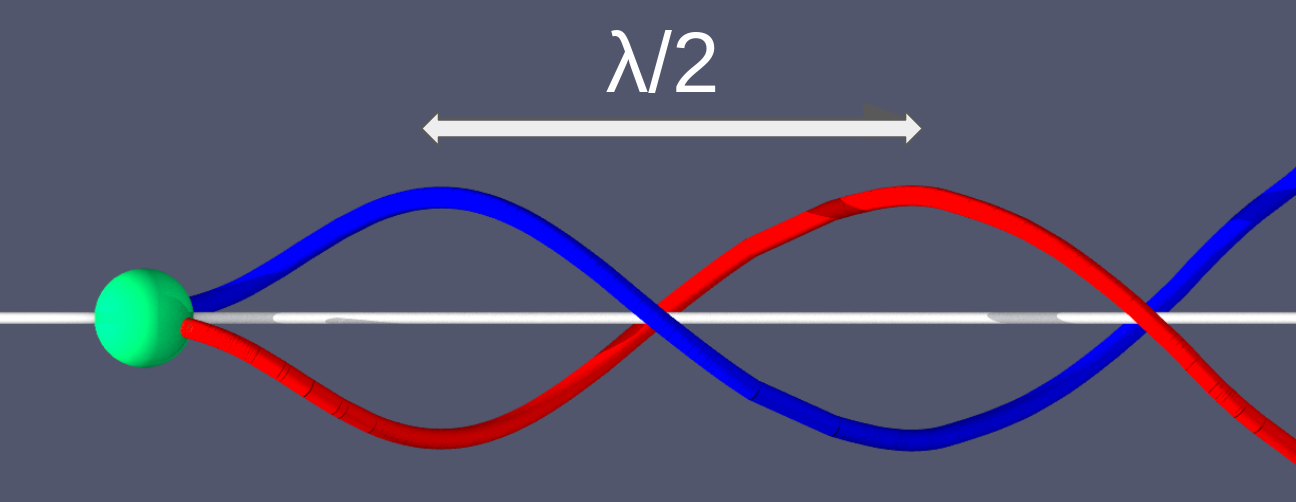}
        }}}
        \caption{Time asymptotic position of a generic magnetic field line is a double helix with helical pitch $\lambda$. Colour-code refers to the sign of $B_z$.} 
        \label{fig_7}
\end{figure}

As already mentioned, numerical simulations show that the magnetic tail of Uranus at solstice is spatially twisted downstream of the planet (along the $z$ axis in our model). As distance from the planet grows, the twist along the $z$ axis rapidly converges towards a sharply defined spatial oscillation scale $\lambda=2\pi/k$. In simulations, the spatial oscillation may be best made evident by plotting an 'old' field line, meaning a field line that had time to become twisted several times with respect to the position of its foot points on the planet. Such a field line is located close to the centre of the tornado (showing up as vortices in the cuts of Figures \ref{fig_5} and \ref{fig_6}) over most of its length. An example of such a field line (the same as the grey line in Fig. \ref{fig_4}) is shown in Figure \ref{fig_7}.  
It is probably illusory to search for an expression of the wavelength $\lambda$ as a function of the global parameters of the problem. On one side, even while keeping within the strict limits of MHD theory and for the simple configuration of Figure \ref{fig_1}, the number of parameters required to fully specify the problem is pretty large as it includes the wind parameters (at least velocity, density, and pressure) as well as the planetary parameters (at least the dipole field strength, the rotation speed, and most likely the planetary radius). In addition, $\lambda$ may also depend on the conditions at the planetary surface such as the ionospheric resistivity and the allowed mass flux from or towards the planet.
Expressing $\lambda$ as a function of the global parameters of the problem may be difficult or impossible, but expressing $\lambda$ as a function of the plasma parameters in the tail or in some specific parts of is clearly conceivable. In this respect, we already mentioned in Section \ref{sec_perp_z} that the $z$ component of the plasma velocity near the field line turning point is equal to the phase speed of the whole magnetic structure. Thus, a measure of the velocity $v_z$ of the turning point of a field line allows us to write $\lambda=\tau v_z$ where $\tau=2\pi/\Omega$ is the planet's rotation period. 
In the next section, we show that $\lambda$ can also be expressed as a function of the Alfvén speed measured near the centres of the vortices described earlier.

\subsection{Equilibrium of forces near the vortices' centres}

\begin{figure}
        \center{\rotatebox{0}{\resizebox{0.45\textwidth}{!}{\includegraphics{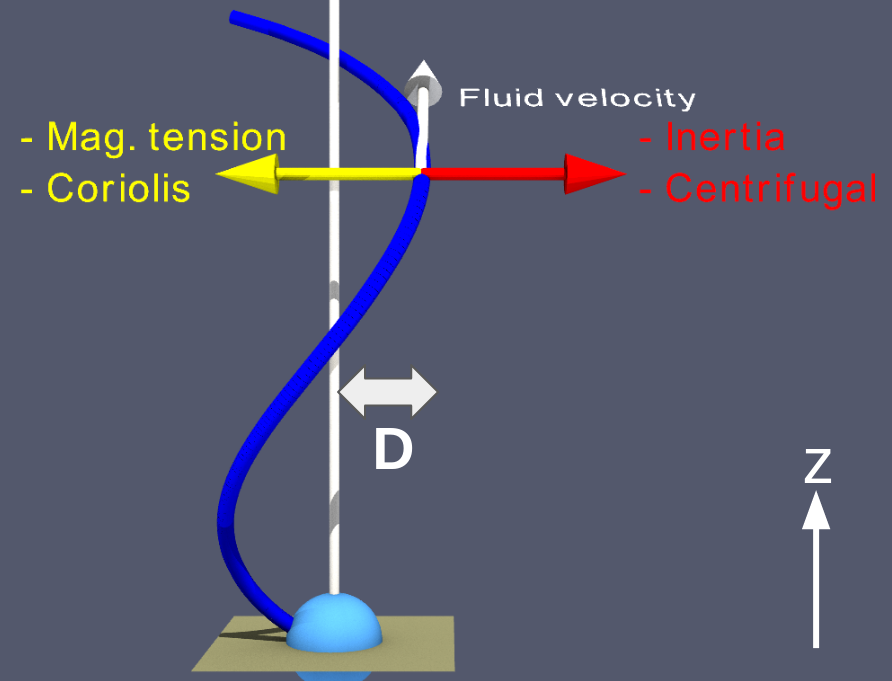}
        }}}
        \caption{In the planetary (rotating) frame, field lines converge towards an asymptotic position where they become motionless and where the fluid velocity and field lines are aligned. It turns out that only the four mentioned terms of the  momentum equation (see Eq. (\ref{eq_momentum_2})) ensure the equilibrium.} 
        \label{fig_8}
\end{figure}

We now know that in the frame of our simplified configuration of the Uranus-at-solstice magnetosphere, all field lines migrate towards the same asymptotic position reminiscent of the eye of an atmospheric tornado. In Figure \ref{fig_8}, a portion of the tornado's eye is  represented as a tube. 
Since in the rotating frame the system can be assumed to be in a quasi-steady state, we may write the momentum equation as follows: 
\begin{equation}
\rho {\bf v}\cdot{\nabla \bf v}=-\nabla p_{\rm tot}+{\bf B}\cdot{\nabla \bf B} + 
2\rho{\bf v}\times{\bf \Omega} +\rho{\bf D}\Omega^2\label{eq_momentum}
,\end{equation}
where $\rho$ is the plasma mass density, $p_{\rm tot}\equiv p+B^2/2$ the total (plasma+magnetic) pressure, and $\bf D$ the vector pointing from the $z$ axis to the point under consideration. To avoid confusion, we name all terms in Eq. (\ref{eq_momentum}). On the left hand side, we have the inertial term, while on the right hand side, we have, successively, the pressure gradient term, the magnetic tension term, the Coriolis term, and the centrifugal term. Also, in order to ease readability, we set the vacuum permeability to $\mu_0=1$ in Eq. (\ref{eq_momentum}). Unless specified otherwise, we adhere to this convention for the remainder of the paper. 
Equation (\ref{eq_momentum}) can be simplified further near the vortex centre where the total pressure is expected to have an extremum (as for the gas pressure in the centre of a cyclone), so one may neglect the pressure gradient term in Eq. (\ref{eq_momentum}). In addition, as illustrated in Figures \ref{fig_5} and \ref{fig_6}, field lines become static when approaching the asymptotic positions implying that the fluid velocity must become more and more aligned on the magnetic field. The asymptotic position being characterised by a wavelength $\lambda=2\pi/k$ in the $z$ direction and an excursion $D$ with respect to the $z$ axis we can then write 
\begin{equation}
\frac{B_\perp}{B_z}=\frac{v_\perp}{v_z}=Dk\label{eq_bperp_over_bz}
,\end{equation}
where the direction $\perp$ has to be considered with respect to the $z$ axis. 
Accordingly, equation (\ref{eq_momentum}) can be simplified and rewritten as  
\begin{equation}
-\rho k v_z v_\perp = -k B_z B_\perp -2\rho v_\perp \Omega 
+\rho D \Omega^2\label{eq_momentum_2}
,\end{equation}
an equilibrium graphically illustrated in Figure \ref{fig_8}. 
Equation (\ref{eq_momentum_2}) has a surprisingly simple general solution: 
\begin{equation}
\lambda=\tau(v_z\pm B_z/\sqrt{\rho})\label{eq_lambda}
,\end{equation} 
where $B_z/\sqrt{\rho}$ is the Alfvén speed in the $z$ direction (not along the field direction). Expression (\ref{eq_lambda}) with the $+$ sign was tested by \cite{Toth_etal_2004}, who found a significant discrepancy between and estimate of the right hand side of Expression (\ref{eq_lambda}), giving $\lambda \approx 1300\, R_{\rm U}$ 
and the wavelength observed in their 
simulation $\lambda_{\rm obs}\approx 900\,R_{\rm U}$. The authors ascribe the discrepancy to a frictional interaction between the solar wind plasma and the helical tail. Such an effect cannot be excluded. However, besides the fact that the tailward extension of their simulation domain was too small ($=704\,R_{\rm U}$) to contain one full wavelength $\lambda,$ the other important point we raise here is that, strictly speaking, Eq. (\ref{eq_lambda}) is only valid at two particular positions in the $(x,y)$ plane, that is, at the centre of the vortices in Figure \ref{fig_6}. Instead, \cite{Toth_etal_2004} evaluated the right-hand side of Eq. (\ref{eq_lambda}) by taking averages  of $v_z$, $B_z,$ and $\rho$ across the magnetic tail. Now, all simulations, including the ones by \cite{Toth_etal_2004} show that the plasma parameters vary considerably in the $(x,y)$ plane, even at large distances from the planet.

\section{Order of magnitude estimate of the currents induced by the rotation\label{sec_current}}
The twisting and stretching of the magnetic field lines in the tail do generate currents flowing to or from the polar regions in order to close by flowing over the ionosphere. Taking $\overline{j}_z \approx \overline{B}_\perp/D$ as an estimate of the average $z$ component of the current at a position, $z,$ down the tail, where the torsion of the field lines by the rotation is effective (e.g. at the position of the first vortex in Figure \ref{fig_5}), we find (using (\ref{eq_bperp_over_bz}) as an estimate for the average $\overline{B}_z$ field) that $\overline{j}_z\approx k\overline{B}_z$. Since all the magnetic field lines crossing  half of the surface delimited by the magnetopause in the $(x,y)$ plane of Figure \ref{fig_6} (of order $\sim \pi D^2$) stem from an area $S$ surrounding one of the magnetic poles, it follows that $\overline{j}_z\approx k S B_p/(\pi D^2),$ where $B_p$ is the surface field strength given by $B_p=2M/R_{\rm U}^3$ and $M$ being the planetary dipole strength (units ${\rm T\,m}^3$). The total current flowing through a polar region is then $I_z\approx \pi D^2\,\overline{j}_z\approx kSB_p,$ where $SB_p$ should be interpreted as the magnetic flux through the planetary surface due to all field lines from one hemisphere being sufficiently twisted to contribute to the current. For Uranus, $M=3.9\,10^{17}\,{\rm Tm}^3,$ and simulations indicate that $\lambda=2\pi/k \approx 10^3 R_{\rm U}$. The total current at one pole can thus be estimated to be of the order $I_z\approx (S/\pi R_{\rm U}^2)\times 1.9\,10^7 {\rm A} $ with a surface ratio $(S/\pi R_{\rm U}^2),$ which we may arbitrarily guess to fall within the range of $10^{-1}$ to $10^{-3}$.       

\section{Limits of the model\label{sec_limits}}
The (inevitable) formation of magnetic vortices anchored on the planet is a challenging aspect of the model. Indeed, a continuous injection of magnetic field lines at the equator and a continuous concentration of field lines at the centre of the vortices is incompatible with the assumption of time steadiness given that the field intensity in the vortices increases with time. One may  safely argue that the incoherence of the model is more conceptual than practical, at least if the characteristic timescale for the flux concentration is long compared to the 
rotation period of the planet. A steady-state Dungey-type cycle \citep{Dungey_1961}, where magnetic flux emerges from the dayside planetary surface and disappears through the nightside surface of the planet, could be invoked. The problem is that the Dungey mechanism requires reconnection through the magnetopause and may only be a solution if reconnection impedes the twisting of the magnetic field lines. Such an eventuality (not allowed in our model, where the interplanetary plasma does not carry any magnetic field) cannot definitely be excluded for Uranus, as the duration of the Dungey cycle is comparable to the rotation period of the planet \cite[e.g.][]{Bagenal_2013}. If, however, as simulations seem to suggest, the twisting of the magnetic field lines is real, even for the relatively slowly rotating Uranus, another mechanism must impede  the vortical concentration of magnetic flux. The difficulty stems from the fact that there is no evident way to limit the vortical concentration of the magnetic flux other than by stopping the emergence of magnetic flux, in which case the only theoretically possible system would be a static one. On the other hand, if one assumes (1) a rigorous steady state; (2) that magnetic flux emerges through the planetary surface; and (3) the formation of a twisted magnetic tail, then, a mechanism of evacuation of the magnetic flux from the vortex centres must necessarily operate. 
Such a mechanism implies the topological reconfiguration of the stretched magnetic field lines approaching the vortex centre to detach them from the planet. However, a topological reconfiguration of a field line such as the one in Figure \ref{fig_7} is not easily imaginable in a globally steady state system. Even a hypothetical reconfiguration through reconnection with an outer magnetic field through the magnetopause seems unrealistic given that the asymptotic field lines are located deep inside the volume delimited by the magnetopause (see Figures \ref{fig_5} and  \ref{fig_6}). A reconfiguration through self-reconnection appears unrealistic for the same reasons, portions of a field line approaching the asymptotic position with opposite $B_z$ polarity being always separated by a large distance (in comparison to an ion's Larmor radius) either of order $2D$ in the tail (see Figure \ref{fig_8}) or of order $R_{\rm U}$, near the magnetic foot points. Finally, despite appearances, and the high degree of symmetry, the system is a complex one, plagued with ill-defined boundaries, for which no stability analysis has ever been made, even within the restrictive frame of ideal MHD. In addition, since the magnetic structure is  neither axisymmetric nor periodic, we expect its global stability properties to differ substantially from the well-documented stability properties of tokamak plasmas. One may consider that the system is approximately axisymmetric in the vicinity of the vortex centre (where magnetic flux slowly concentrates) with a component of the flow velocity perpendicular to the axis of the vortex. Under such circumstances, \cite{Goedbloed_etal_2004} showed that the axisymmetric structure turns unstable as soon as the flow velocity exceeds the slow mode velocity. Whether or not this 'trans-slow' instability plays a role in the reconfiguration process of a complex double helix magnetic tail cannot be established without a more in-depth numerical or theoretical stability analysis. It should be considered merely as one out of many other possible ideal (and non-ideal) fluid instabilities that may affect the system on the long timescale associated with the concentration of magnetic flux near the asymptotic position.   

The conclusion we draw at this point is that the model is not compatible with a rigorous steady state. Catastrophic reconfigurations on a yet undetermined timescale must be invoked. Successive reconfigurations likely occur on a timescale of the order of the time necessary to substantially increase the vortical magnetic flux, but other external factors such as variations of the solar wind dynamic pressure or of the interplanetary magnetic field orientation (not considered in the model) may also play a role in triggering the reconfiguration. 

\section{Speculative extrapolation to the general case\label{sec_speculations}}

Our model is admittedly an oversimplification for Uranus. 
With regard to what we learned about the planet after Voyager's flyby in January 1986 \citep[see][]{Ness_etal_1986}, the most drastic simplification is the assumption of a $90circ$ angle between the planetary rotation axis and its  magnetic dipole. While the consequences of having reduced the angle between the rotation axis and the solar wind direction from $7^\circ$ to $0^\circ$ are certainly minor, the consequences of having increased the angle between the rotation axis and the planetary magnetic field axis from $59^\circ$ to $90^\circ$ must be considered. Clearly, the principal difference of the real configuration with respect to the one in Figure \ref{fig_1} is that with a $59^\circ$ the two magnetic poles are not equivalent in the real case, where one magnetic pole is always on the dayside, while the other is always on the nightside. The system is still time stationary in the rotating frame, at least on not-too-long timescales, as discussed in Section \ref{sec_limits}, and as long as no interplanetary field is assumed, but many of the other symmetries exposed in Section \ref{sec_symmetries} are lost. 
The simulations by \cite{Toth_etal_2004} who used the effective orientations (see their Figure 17) do indeed show that the symmetry between the polarities of the stretched and twisted field lines is broken, especially in the vicinity of the planet. The main reason is that the field lines emanating from the nightside pole have direct access to the volume enclosed by the magnetopause downstream of the planet, while the field lines emanating from the dayside pole have to circumvent the planet first and therefore tend to run along the magnetopause. Interestingly, their simulation also shows that the asymmetry fades away rapidly downstream (over a distance much shorter than $\lambda$), suggesting that the spiralling movement of the magnetic field lines observed in our model may not apply to the field lines emanating from the nightside pole but also (to some extent) to the field lines emanating from the dayside pole. Whether the spiralling motion drives the magnetic field lines towards a curve as in our symmetric model (cf. Figure \ref{fig_8}), or much more likely towards a surface, is a question that may be investigated in new dedicated simulations. 
   
The other important simplification of our model is the unmagnetised wind assumption, which excludes the possibility of magnetic reconnection at the magnetopause. Magnetic reconnection has been observed in simulations of a fast-rotating Uranus-type magnetosphere in the same configuration as the one of Figure \ref{fig_1} except for the presence of a magnetised wind  \cite[see][Figure 1]{Griton_etal_2018}.
The consequence of reconnection in the sketches of Figures \ref{fig_5} and \ref{fig_6} would be an 'erosion' of the outermost magnetic surfaces located near the magnetopause. However, a complete reconnection-driven disruption of all magnetic surfaces surrounding the asymptotic position is much too slow a process to be effective over distances of order $\lambda$ downstream of the planet. For example, no significant erosion was observed in \cite{Griton_etal_2018}. 

\section{Conclusions\label{sec_conclusions}}
We present an MHD-based physical model for the magnetic tail of a Uranus-type planet at solstice time. In order to limit complexity and facilitate understanding, we based our model on a symmetrised version of the Uranian system at solstice plunged in a unmagnetised solar wind. Consequently, in our model, the planet's rotation axis is parallel with respect to the solar wind direction, the (centred) planetary magnetic dipole is perpendicular to the rotation axis, and no magnetic reconnection is possible at the magnetopause. Given these assumptions, a quasi-steady state can be reached in the rotating planetary frame. The main consequences we draw from the model are as follows: (1) magnetic field lines emerge near the magnetic equator, and  their foot points move on a spiralling path (see Figure \ref{fig_5}) towards a final point located nightside of the nominal magnetic poles; (2) in the planetary (rotating) frame, each spiralling field line describes tornado-shaped magnetic surfaces (see Figure \ref{fig_4});  (3) the tornados' eyes are thus surrounded by a \textit{\emph{'millefeuille'}} of magnetic surfaces, one for each emergence point on the planetary surface; (4) the two eyes are interlaced curves extending downstream of the planet with a characteristic wavelength $\lambda$ given by Eq. (\ref{eq_lambda}). We note that the model implies a continuous accumulation of magnetic flux in the eyes and is thus incompatible with a rigorous steady state. Therefore, in the long-term, the system is expected to turn unstable and possibly subject to catastrophic reconfiguration of its magnetic structure, even under constant wind conditions.   

\begin{acknowledgements} Fruitful discussions with Camille Noûs, from the newly founded Laboratoire Cogitamus, must be warmly acknowledged in this place. Thank you to the referee for his/her highly pertinent and constructive comments and suggestions.      
\end{acknowledgements}  

\bibliographystyle{aa}
\bibliography{uranus_solstice_model.bib}

\end{document}